\documentclass[lettersize,journal]{IEEEtran}

\usepackage{amsmath}
\usepackage{amssymb,amsthm}
\usepackage{graphicx}
\usepackage{psfrag}
\usepackage{epsfig}
\usepackage{dblfloatfix}
\usepackage{kotex}
\usepackage{color}
\usepackage{latexsym, amssymb, array, multirow}
\usepackage{algorithm, algpseudocode}
\usepackage{subcaption}
\usepackage{booktabs}

\makeatletter
\def\BState{\State\hskip-\ALG@thistlm}
\makeatother

\theoremstyle{definition}

\theoremstyle{remark}

\newcommand{\sr}{{\rm S\!R}}
\newcommand{\sd}{{\rm S\!D}}
\newcommand{\rd}{{\rm R\!D}}

\newcommand{\st}{{\rm S}}
\newcommand{\rt}{{\rm R}}

\newcommand{\CN}{\mathcal{CN}}

\newcommand{\Complex}{\mathbb{C}}

\begin{document}
\title{Efficient Hybrid Amplitude-Phase Quantization for Multi-Antenna Relay System}
\author{Changdae Kim, \IEEEmembership{Student Member,~IEEE}, and Xianglan Jin, \IEEEmembership{Member,~IEEE}
\thanks{

The authors are  with the Division of Electronic Engineering and IT Convergence Research Center, Jeonbuk
National University, Jeonju 54896, Korea (email: jinxl77@jbnu.ac.kr).
}
}

\maketitle

\begin{abstract}
This letter explores relay quantization in multi-antenna quantize–forward (QF) relay systems. 
Existing methods, such as uniform phase quantization (U-PQ) and uniform amplitude-phase quantization (U-APQ), suffer from performance saturation and high memory demands. 
To overcome these limitations, we propose hybrid amplitude-phase quantization (H-APQ), which adaptively quantizes received signal amplitudes based on their relative magnitudes while applying uniform quantization to individual phases. 
H-APQ significantly reduces memory consumption at the relay while maintaining strong overall performance, offering an efficient solution for multiple-input multiple-output (MIMO) QF relay systems.
\end{abstract}


\IEEEpeerreviewmaketitle

\section{Introduction}\label{sec:intro}

 Relay-assisted cooperative communication enhances wireless system performance by introducing an intermediate relay between a transmitter (source) and a receiver (destination) in a point-to-point (P2P) communication link. 
 Traditional relay strategies include amplify--forward (AF) \cite{cooperative-diversity2004} and decode--forward (DF) \cite{cooperative-diversity2004,relay-df2011}. 
 However, these methods have inherent limitations: AF relaying requires substantial memory to store continuous received signals, while DF relaying imposes high computational complexity due to the need for decoding and re-encoding. Additionally, both approaches rely on accurate channel estimation, further increasing system overhead.

Quantize--forward (QF) relaying provides a practical alternative by applying quantization to the received signals, significantly reducing memory requirements \cite{QF2008,JinQFtvt2022,JinQFtwr2023, AQF2006,JinAQFaccess2022,JinMIMOAQF_TVT2024}. 
Unlike AF and DF, QF does not require channel state information (CSI), making it well-suited for real-world communication scenarios with limited feedback. Moreover, it effectively addresses the drawbacks of AF and DF by reducing memory usage and computational complexity.

The most widely used quantization scheme in QF relay systems is uniform phase quantization (U-PQ), which quantizes and forwards only the phase information of the received signals \cite{QF2008,JinQFtvt2022}. 
However, in multiple-input multiple-output (MIMO) QF relay systems, the performance of U-PQ saturates beyond a certain number of quantization bits \cite{AE-MIMO-QF2024}. 
This occurs because U-PQ accounts only for phase relationships between antennas while disregarding amplitude variations. 
To overcome this limitation, the uniform amplitude-phase quantization (U-APQ) algorithm was introduced in autoencoder-based relay systems \cite{AE-MIMO-QF2024}. 
By incorporating both uniform phase quantization and uniform amplitude quantization, U-APQ outperforms U-PQ when using the same number of quantization bits. 
However, two key challenges remain: the amplitudes of the received signals are distributed over the range $(0,+\infty)$, requiring normalization prior to quantization. 
Additionally, because the normalized amplitudes are non-uniformly distributed within $(0,1)$, with denser clustering in specific regions, they must be represented using a sufficiently large number of bits.

To address these issues, this letter proposes a novel hybrid amplitude-phase quantization (H-APQ) algorithm. 
While maintaining the same uniform phase quantization as U-APQ, H-APQ introduces a new amplitude quantization method, ordered amplitude quantization (O-AQ), which assigns quantization levels based on the relative magnitudes of received signal amplitudes. 
The proposed approach significantly reduces memory requirements while achieving comparable performance, making it an efficient solution for MIMO QF relay systems.

This letter is structured as follows. Section \ref{sec:sys_model} describes the autoencoder-based QF relay system and reviews the existing U-PQ and U-APQ algorithms, followed by the introduction of the proposed H-APQ method. Section \ref{sec:numerical_results} presents numerical results demonstrating the effectiveness of H-APQ. Finally, Section \ref{sec:conclusion} concludes the letter.
\begin{figure}
\centering\includegraphics[width=1.0\linewidth]{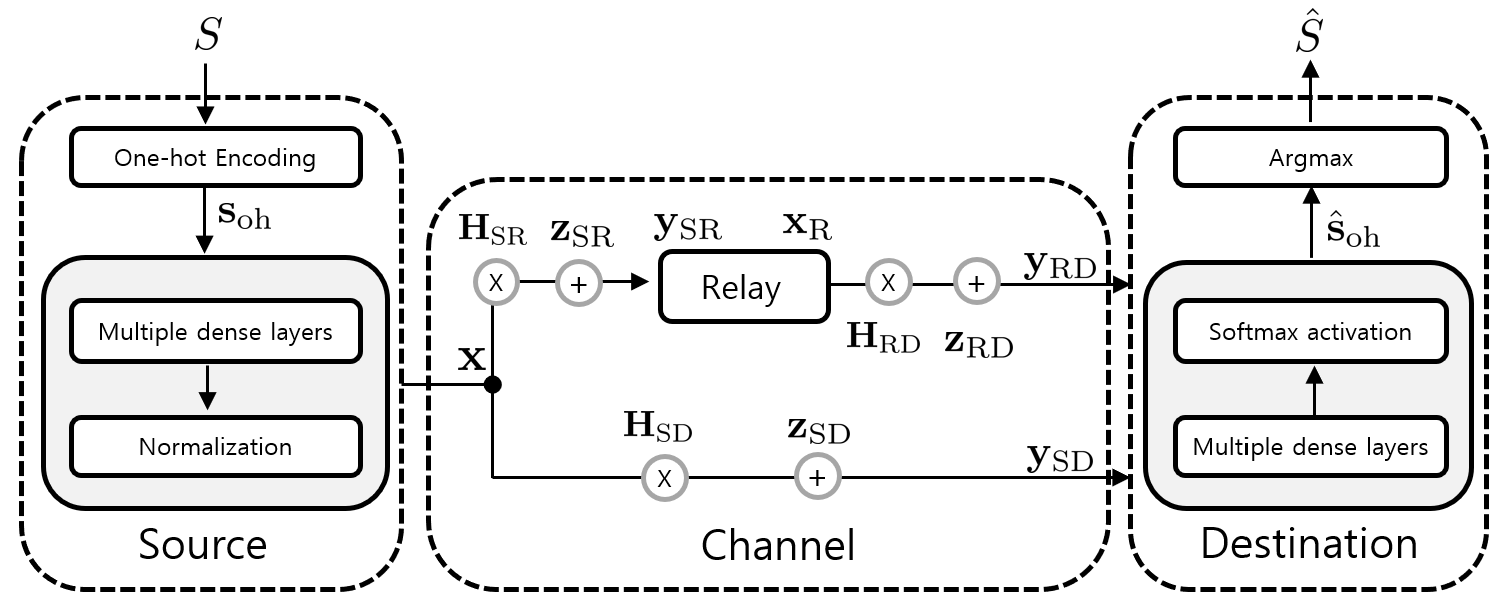}
\caption{{\label{fig:qfrelay}{An autoencoder-based QF relay   system}.}}
\end{figure}

\section{Autoencoder- Based QF Relay System}\label{sec:sys_model}
As illustrated in Fig. \ref{fig:qfrelay}, we consider an autoencoder-based MIMO QF relay communication system\footnote{The proposed H-APQ can be applied to traditional communication systems that include channel encoding, modulation, demodulation, and decoding. However, we focus on an autoencoder-based relay system because the existing U-APQ method, to which H-APQ is compared, is designed specifically for autoencoder-based relay systems and cannot be applied in traditional communication frameworks. 
}. 
The autoencoder model comprises an encoder, a decoder, and a channel. In this relay cooperative communication system, the source functions as the encoder, while the destination serves as the decoder. 
The communication channel between the encoder and decoder includes the source-destination (SD) link, representing the direct path, and the source-relay-destination (SRD) link, which involves the relay.

In this section, we first describe the quantized relay channel, review existing quantization algorithms, and introduce a novel quantization algorithm. 
Finally, we provide a brief overview of the autoencoder-based relay system.
 
\subsection{QF relay channel model}\label{sec:qf_ch}
A relay cooperative communication system basically transmits a message from the source to the destination through the cooperation of the relay.
The source, destination, and relay are equipped with $N_{\mathrm S}, N_{\mathrm D}$ and $N_{\mathrm R}$ antennas, respectively.

In the first time slot, the source broadcasts   symbols   $ {\bf x} = [x_1 \ x_2 \ \cdots \ x_{N_{\mathrm S}}]^{T} \in\Complex^{{N}_{\mathrm S}}$ with $\|{\bf x}\|^2 = 1$  to the relay and destination.
Then, the relay quantizes the received signals and stores the quantized information in memory.
 
 In the second time slot where the source is in a waiting state, the relay generates new signals ${\bf x}_{\rt}$ with $\|{\bf x}_{\mathrm R}\|^2 = 1$ and transmits them to the destination after reloading the quantized information from the memory.
 
 As a result, the received signals at the relay and destination for the first and the second time slots can be written
\begin{align}
	{\bf y}_{\sr} &= {\bf H}_{\sr}{\bf x} +{\bf z}_{\sr}\label{eq:sr}\\[-2pt]
	{\bf y}_{\sd} &= {\bf H}_{\sd}{\bf x} +{\bf z}_{\sd} \label{eq:sd} 
\end{align}
and
\begin{align}
	{\bf y}_{\rd} &={\bf H}_{\rd}{\bf x}_{\rt}+{\bf z}_{\rd}\label{eq:rd}
\end{align}
respectively, where ${\bf H}_{\sr}\in\Complex^{N_{\mathrm R} \times N_{\mathrm S}},{\bf H}_{\sd}\in\Complex^{N_{\mathrm D} \times N_{\mathrm S}}$ and ${\bf H}_{\rd}\in\Complex^{N_{\mathrm D} \times N_{\mathrm R}}$  represent the channel coefficient matrices for the source-relay (SR), SD, and relay-destination (RD) links, respectively.
  ${\bf z}_{kl} \sim\CN(0,\sigma^2I_{N_l}), k \in\{ {\mathrm S}, {\mathrm R} \},l \in\{ {\mathrm R}, {\mathrm D} \}$ are circularly symmetric complex
Gaussian distributed noise vectors. 

Then the transmission signal-to-noise ratio (SNR) for all links is  $1/\sigma^2$.
\begin{figure}[t]
    \centering
    \begin{subfigure}[t]{0.8\linewidth}
        \centering
        \includegraphics[width=\linewidth]{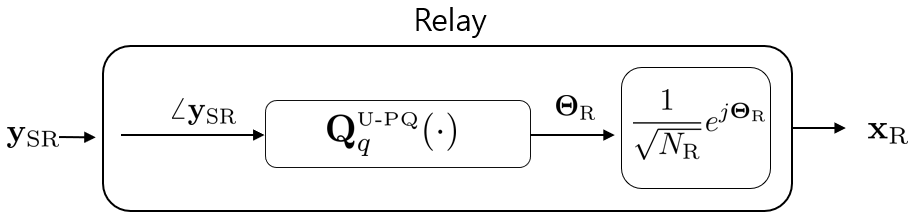}
        \caption{U-PQ\\[10pt]}
        \label{fig:U-PQ}
    \end{subfigure}
 
    \begin{subfigure}[t]{0.8\linewidth}
        \centering
        \includegraphics[width=\linewidth]{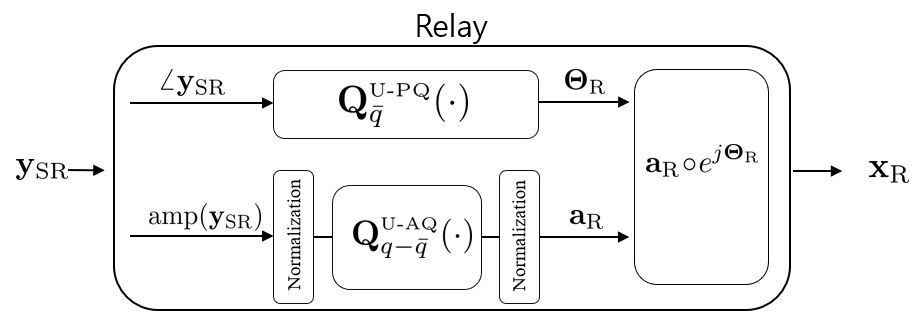}
        \caption{U-APQ\\[10pt]}
        \label{fig:U-APQ}
    \end{subfigure}
    \hfill
    \begin{subfigure}[t]{0.8\linewidth}
        \centering
        \includegraphics[width=\linewidth]{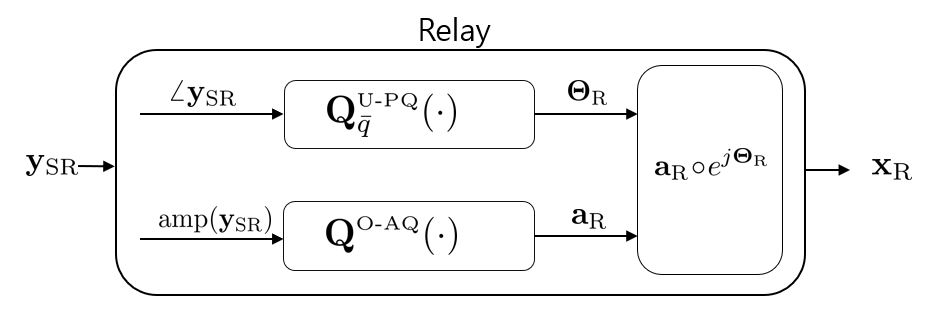}
        \caption{H-APQ\\[10pt]}
        \label{fig:H-APQ}
    \end{subfigure}
    \caption{Quantization algorithms at the relay.}
    \label{fig:total_QF}
\end{figure}

\subsection{Quantization algorithm at relay }\label{sec:qf}

\subsubsection{Uniform phase quantization (U-PQ) \cite{QF2008,AE-MIMO-QF2024}}\label{sec:U-PQ}
 
 As shown in Fig. \ref{fig:U-PQ}, we perform $q$-bit  U-PQ for each phase of the received signals at the relay in the first time slot.
When the phase  of the $i$th received signal satisfies $\frac{(2k-1)\pi}{2^q} < \angle[{\bf y}_{\sr}]_i \leq \frac{(2k+1)\pi}{2^q}$, the quantized phase is obtained as
\begin{align}
	[\boldsymbol{\Theta}_{\mathrm R}]_i = \mathbf{Q}^{\text{\fontsize{6}{6}\selectfont U-PQ}}_{q}(\angle[{\bf y}_{\sr}]_i) = \frac{2\pi k}{2^q}
\end{align}
where $k=0, 1, \ldots, 2^q-1$ and $\mathbf{Q}^{\text{\fontsize{6}{6}\selectfont U-PQ}}_{q}(\cdot)$ is the  element-wise U-PQ.
At this point, the total number of required quantization bits is ${N}_b=qN_{\mathrm R}$ to distinguish all the the possible quantization levels.

 As a result, we have
\begin{align}
	{\bf x}_{\mathrm{R}} = \frac{1}{\sqrt{N_{\mathrm{R}}}} e^{j\boldsymbol{\Theta}_{\mathrm{R}}} = \frac{1}{\sqrt{N_{\mathrm{R}}}} e^{j\mathbf{Q}^{\text{\fontsize{6}{6}\selectfont U-PQ}}_q\left(\angle {\bf {y}}_{\sr} \right)}
\end{align}
where $e^{j\boldsymbol{\Theta}_{\mathrm{R}}} = \begin{bmatrix} e^{j[\boldsymbol{\Theta}_{\mathrm{R}}]_1} & \cdots & e^{j[\boldsymbol{\Theta}_{\mathrm{R}}]_{N_{\mathrm{R}}}} \end{bmatrix}^T$ and each component is divided by $\sqrt{N_{\mathrm{R}}}$ in order to satisfy the power constraint of $\|{\bf x}_{\mathrm R}\|^2 = 1$. 

In the MIMO QF relay system, performance improvement becomes limited  when the number of quantization bits, $q$, exceeds a certain threshold \cite{AE-MIMO-QF2024}. 
This limitation stems from the difficulty in accurately forwarding the amplitude information of the received signals.

\subsubsection{Uniform amplitude-phase quantization (U-APQ) \cite{AE-MIMO-QF2024}}\label{sec:U-APQ}

 As shown in  Fig. \ref{fig:U-APQ},  the U-APQ algorithm  quantizes both the amplitude and phase of the received signal at the relay with  the same total number of bits as U-PQ.
Among the $q$ bits, $\bar {q}$ bits are allocated to phase quantization denoted as $\mathbf{Q}^{\text{\fontsize{6}{6}\selectfont U-PQ}}_{\bar{q}}(\cdot)$, and $q-\bar {q}$ bits are allocated to amplitude quantization.
Therefore, the total number of quantization bits is ${N}_b=qN_{\mathrm R}$.

Since the amplitudes of the received signals are distributed over the range $(0,+\infty)$,  the amplitudes of the $N_{\mathrm R}$ signals are first normalized.
Then $q-\bar {q}$ bit uniform amplitude quantization (U-AQ),  $\mathbf{Q}^{\text{\fontsize{6}{6}\selectfont U-AQ}}_{q-\bar{q}}(\cdot)$, is applied in the range of $(0,1)$.
After that, the normalization process is running again to satisfy the power constraint of $\|{\bf x}_{\mathrm R}\|^2 = 1$.
Finally, by combining the quantization information for the phases and amplitudes,   the relay symbols vector ${\bf x}_{\mathrm R}$ is  generated as
\begin{align}\label{eq:apq_signal}
{\bf x}_{\mathrm R} = {\bf a}_{\mathrm R}\circ e^{j\boldsymbol{\Theta}_{\mathrm{R}}}
\end{align}
where $\circ $ is an element-wise multiplication operation of vectors, $\boldsymbol{\Theta}_{\mathrm R} = \mathbf{Q}^{\text{\fontsize{6}{6}\selectfont U-PQ}}_{\bar {q}}(\angle{\bf y}_{\sr})$ is the result of phase quantization.
The vector ${\bf a}_{\mathrm R}  $ is the final output of the amplitude component after normalizing the uniformly quantized amplitudes, $ \mathbf{Q}^{\text{\fontsize{6}{6}\selectfont U-AQ}}_{q-\bar {q}}\left( \frac{{\rm amp}({\bf y}_{\sr})}{\|{\bf y}_{\sr}\|} \right)$.
Here ${\rm amp}({\bf y}_{\sr})=[\alpha_1~\cdots~\alpha_{N_{\rm R}}]^T$ is a vector containing  the amplitudes of $N_{\rm R}$ received signals.

Due to the non-uniform distribution of normalized amplitudes within $(0,1)$, with higher density in certain regions, distinguishing amplitudes becomes challenging when fewer memory bits are used.
\begin{figure}
\centering\includegraphics[width=0.9\linewidth]{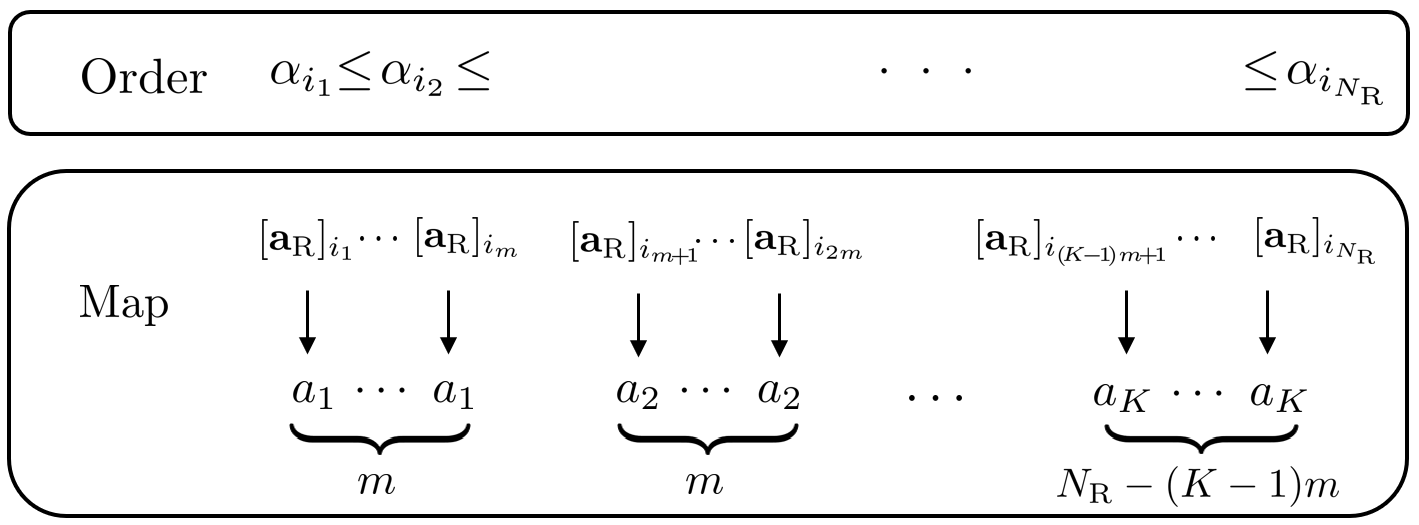}
\caption{Ordered amplitude quantization $\mathbf{Q}^{\text{\selectfont O-AQ}} (\cdot)$.}
\label{fig:oaq}
\end{figure}

\subsubsection{Novel hybrid amplitude-phase quantization (H-APQ)}\label{sec:H-APQ}

 To better distinguish amplitudes using fewer memory bits, we propose a H-APQ algorithm, which  proceeds the same  phase quantization  $\mathbf{Q}^{\text{\fontsize{6}{6}\selectfont U-PQ}}_{\bar{q}}(\cdot)$ as   U-APQ, but introduces a novel amplitude quantization method called ordered amplitude quantization (O-AQ), as illustrated in Fig. \ref{fig:H-APQ}.

As depicted in Fig. \ref{fig:oaq}, the O-AQ method first sorts  the received signal amplitudes  in ascending order, yielding the index sequence $\{{i_1}, {i_2},\dots,i_{\!N_{\rm R}}\} $ such that $\alpha_{i_1}\!\!\le\!\!\alpha_{i_2}\!\!\le\!\cdots\!\le\!\alpha_{i_{\!N_{\rt}}}$. 
Then, it assigns quantization levels from the ordered set $\mathcal{A}=\{a_1\!<\!a_2\!<\!\cdots\!<\! a_K\}$, where each level is allocated to $m$ amplitudes, ensuring a structured mapping of signal amplitudes to quantization levels. 
The number of quantization levels is determined as $K=\lceil N_{\rm R} / m \rceil$. 
Larger amplitudes are assigned higher quantization levels, while smaller amplitudes receive lower ones. 
The detailed assignment process is as follows:
\begin{itemize}
\item[--]  The smallest $m$   amplitudes, $\alpha_{i_1},\dots,\alpha_{i_m}$, are assigned the smallest quantization level $a_1$, i.e., \\[-8pt]
  $$[{\bf a}_{\rm R}]_{i_1}~=~ \cdots~=~[{\bf a}_{\rm R}]_{i_m}~=a_1.$$
\item[--] The next $m$ amplitudes $\alpha_{i_{m+1}},\dots,\alpha_{i_{2m}}$ are quantized to $a_2$, i.e., \\[-10pt]
$$ [{\bf a}_{\rm R}]_{i_{m+1}}= \cdots=[{\bf a}_{\rm R}]_{i_{2m}}=a_2.$$ 
\item[--]  This pattern continues  for subsequent groups of $m$ amplitudes.
\item[--]  Finally, the remaining $N_{\rm R}\!-\!(K\!-\!1)m$  largest amplitudes are assigned the highest quantization level $a_K$, i.e., \\[-10pt]
$$ [{\bf a}_{\rm R}]_{i_{(K-1)m+1}}= \cdots=[{\bf a}_{\rm R}]_{i_{N_{\rm R}}}=a_K .$$
\end{itemize}
This   ensures a structured allocation of quantization levels based on the magnitudes of the received signal amplitudes.

Since the $N_{\rm R}$ signals transmitted after quantization must satisfy power constraint of $\|{\bf x}_{\mathrm R}\|^2 = 1$, the components of $\mathcal{A}=\{ a_1, a_2,\ldots, a_K \}$ are subject to
\begin{align}\label{eq:pow1}
	  \sum^{N_{\rm R}}_{i=1} ([{\bf a}_{\rm R}]_i)^2  \!=\! m\sum^{ K\!-\!1 }_{k=1} {a_k}^2 +(N_{\rm R}\!-\!(K\!-\!1)m) {a_K}^2\!=\!1.
\end{align}
If the quantization levels are set as $a_k = k\Delta$ for $k = 1, \dots, K$, the power constraint in \eqref{eq:pow1} determines $\Delta$ as  
\begin{align}\label{eq:pow_delta1}
\Delta = \sqrt{\frac{1}{\frac{m(K-1)K(2K-1)}{6}+(N_{\rm R}-(K-1)m)K^2}}.
\end{align}
This ensures that the quantization set $\mathcal{A} = \{ \Delta, 2\Delta, \dots, K\Delta \}$ satisfies \eqref{eq:pow1}. While this case adopts $a_k = k\Delta$ as the quantization levels, other level selection strategies that meet the power constraint in \eqref{eq:pow1} can also be applied in the O-AQ method.
Thus, unlike  U-APQ, the H-APQ method does not require two normalization steps in the amplitude quantization process.

As a result of H-APQ, ${\bf x}_{\mathrm R}$ can also be written as the form  in \eqref{eq:apq_signal}, 
where  $\boldsymbol{\Theta}_{\mathrm R} = \mathbf{Q}^{\text{\fontsize{6}{6}\selectfont U-PQ}}_{\bar {q}}(\angle{\bf y}_{\sr})$ is the same one  in \eqref{eq:apq_signal}, and ${\bf a}_{\rm R}$ is the result of  O-AQ.

In  O-AQ, the  number  of possible quantized amplitude vectors, ${\bf{a}}_{\rm R}$, is
\begin{align}
& {N_{\rm R}\choose m}{N_{\rm R}-m\choose m} \cdots {N_{\rm R}-(K-1)m\choose N_{\rm R}-(K-1)m}\notag\\[2pt]
& = \frac{N_{\rm R}!}{(N_{\rm R}-(K-1)m)!(m!)^{K-1}}.
\end{align}
Consequently, the number of bits required to distinguish the quantized amplitude information is $\scriptsize  \Big\lceil\! \log_2\!\Big(\!\frac{N_{\rm R}!}{(N_{\rm R}\!-\!(K\!-\!1)m)!(m!)^{K\!-\!1}} \!\Big) \!\Big\rceil$, and this value decreases as $m$ increases.
As a result, the total number of quantization bits required for H-APQ is $ \scriptsize N_b=  \bar {q}N_{\mathrm R} \!+\! \Big\lceil\! \log_2\!\Big(\!\frac{N_{\rm R}!}{(N_{\rm R}\!-\!(K\!-\!1)m)!(m!)^{K\!-\!1}} \!\Big) \!\Big\rceil$.


 Table \ref{tab:bits} provides a summary of the number of quantization bits required for the three quantization methods at the relay.
\begin{table}[h]
    \centering
    \caption{Number of quantization bits.}
    \renewcommand{\arraystretch}{1.5}
    \begin{tabular}{>{\centering\arraybackslash}p{1cm}||>{\centering\arraybackslash}p{0.6cm}|>{\centering\arraybackslash}p{0.77cm}|>{\centering\arraybackslash}p{4.5cm}} 
         &\scriptsize U-PQ &{\scriptsize U-APQ} &\scriptsize H-APQ \\ 
	 \hline \hline
         \# of bits & $qN_{\rm R}$ & {$qN_{\rm R}$} & $\scriptsize \bar {q}N_{\mathrm R} \!+\! \Big\lceil\! \log_2\!\Big(\!\frac{N_{\rm R}!}{(N_{\rm R}\!-\!(K\!-\!1)m)!(m!)^{K\!-\!1}} \!\Big) \!\Big\rceil$
    \end{tabular}
    \label{tab:bits}
\end{table}
\subsection{Autoencoder-based  communication system}\label{sec:autoencoder}

We transmit a message $S\in\mathcal{M}\!=\!\{1,2,\dots, M^{N_{\rm S}}\}$ from the source  to the destination, as shown in Fig. \ref{fig:qfrelay}.
To facilitate this, we employ an end-to-end autoencoder and a submessage one-hot encoding scheme \cite{AE-MIMO-QF2024}.
First, the message $S$ is transformed into a one-hot encoded vector ${\bf s}_{\rm oh}$, which is then processed through multiple dense layers and normalized to produce the signal $\bf x$.
The signal $\bf x$ is transmitted to the QF relay channel using  $N_{\rm S}$ antennas as described in Section \ref{sec:qf_ch}.
At the destination, the received signals, ${\bf y}_{\rm SD}$ and ${\bf y}_{\rm RD}$, are input into a decoder comprising multiple dense layers with softmax activation. 
The decoder outputs an estimate of the one-hot encoded vector, $\hat{\bf s}_{\rm oh}$.

In this autoencoder model, the training process minimizes the categorical cross-entropy loss between the true one-hot encoded  vector ${\bf s}_{\rm oh}$ and its estimation  $\hat{\bf s}_{\rm oh}$.
During the test  phase, the final estimated value $\hat{S}$ is determined as the index of the component with the maximum probability in $\hat{\bf s}_{\rm oh}$.

\begin{figure}[t]
\centering\includegraphics[width=0.8\linewidth]{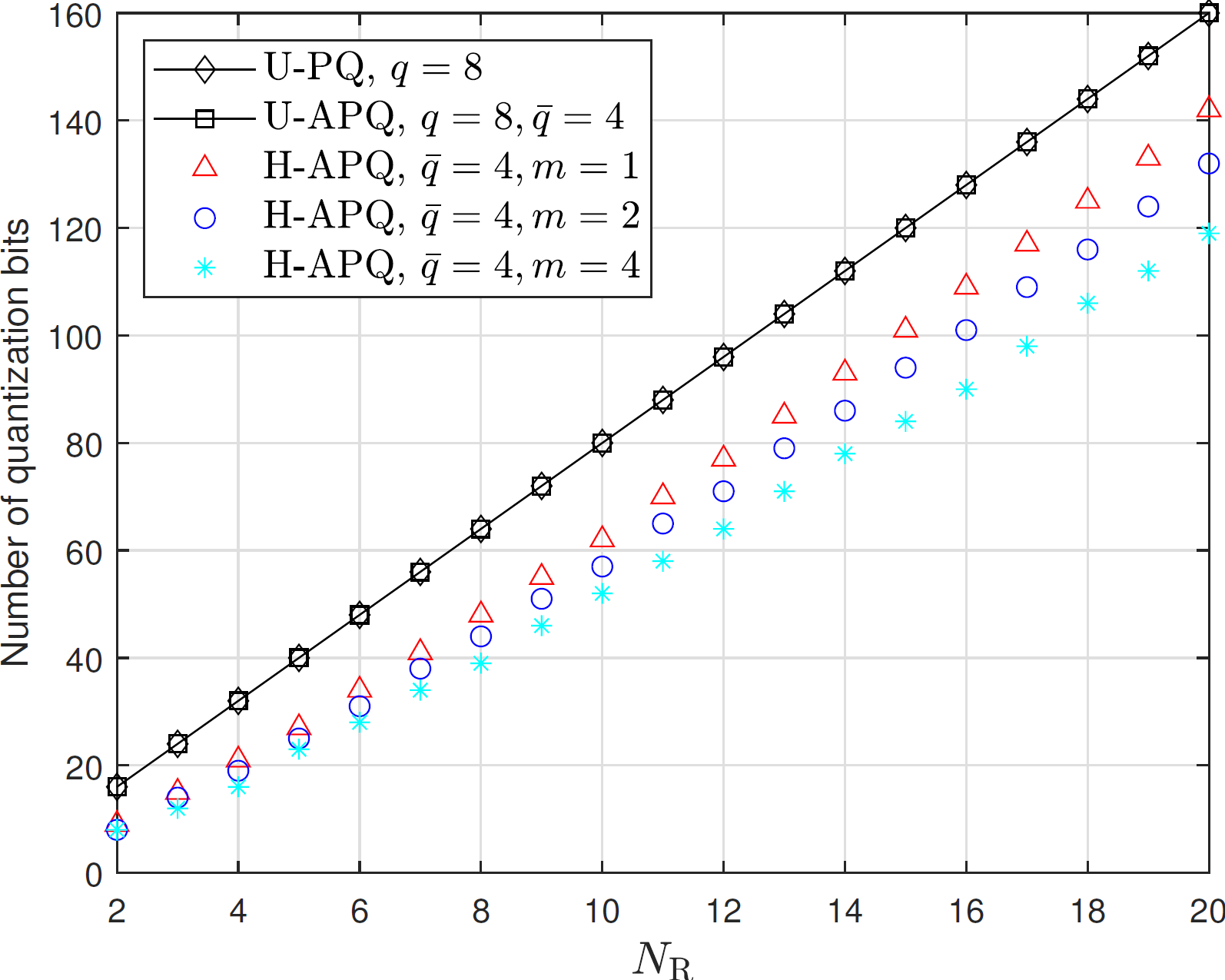}
\caption{Number of quantization bits for various quantization methods at the relay.}
\label{fig:memory_bit}
\end{figure}
 \begin{figure}[t]
    \centering
    \begin{subfigure}{0.8\linewidth}
        \centering
        \includegraphics[width=\linewidth]{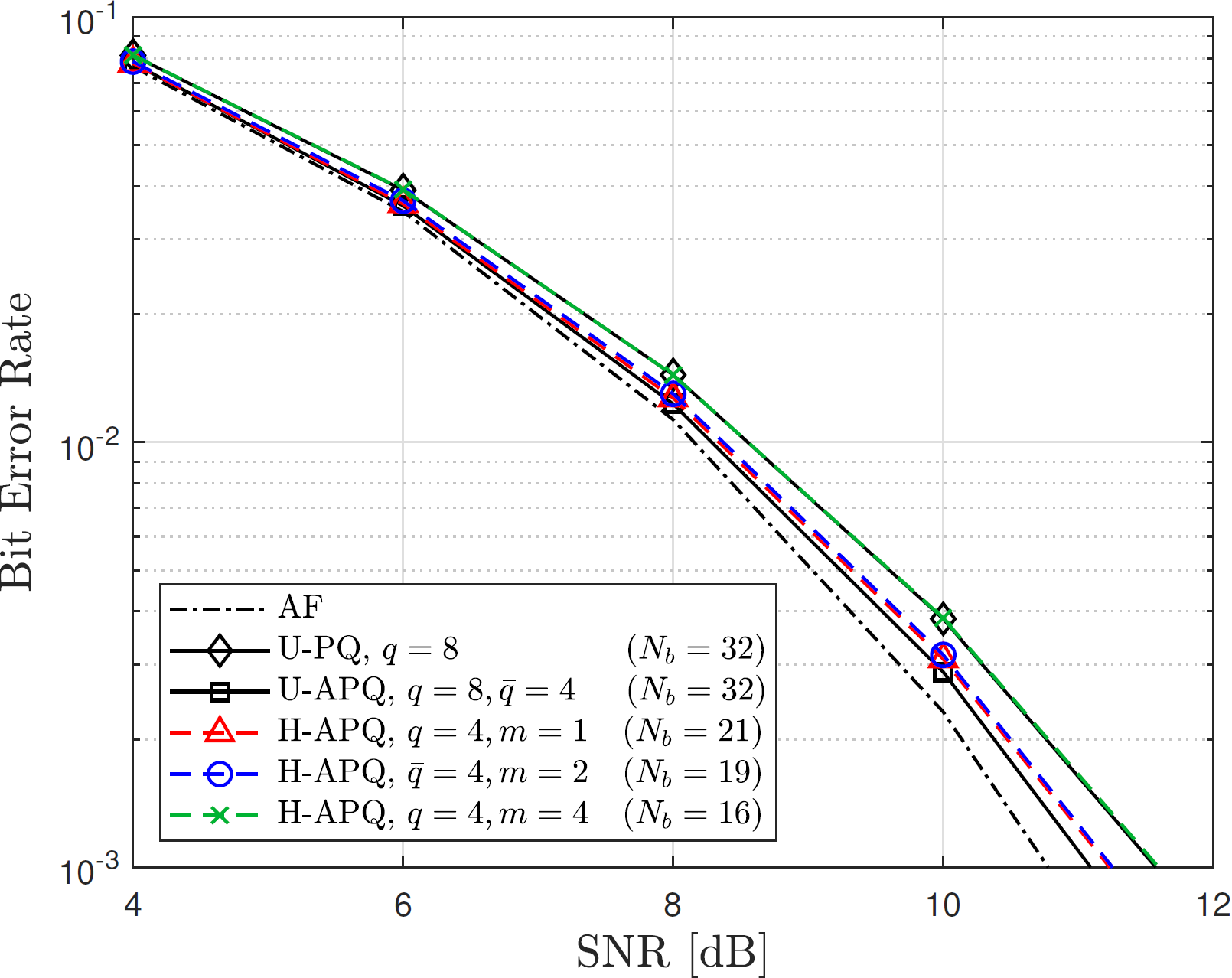}
        \caption{$N_{\rm S}=N_{\rm R}=N_{\rm D}=4$}
        \label{fig:berNR4}
    \end{subfigure} \\[7pt]
    \begin{subfigure}{0.8\linewidth}
        \centering
        \includegraphics[width=\linewidth]{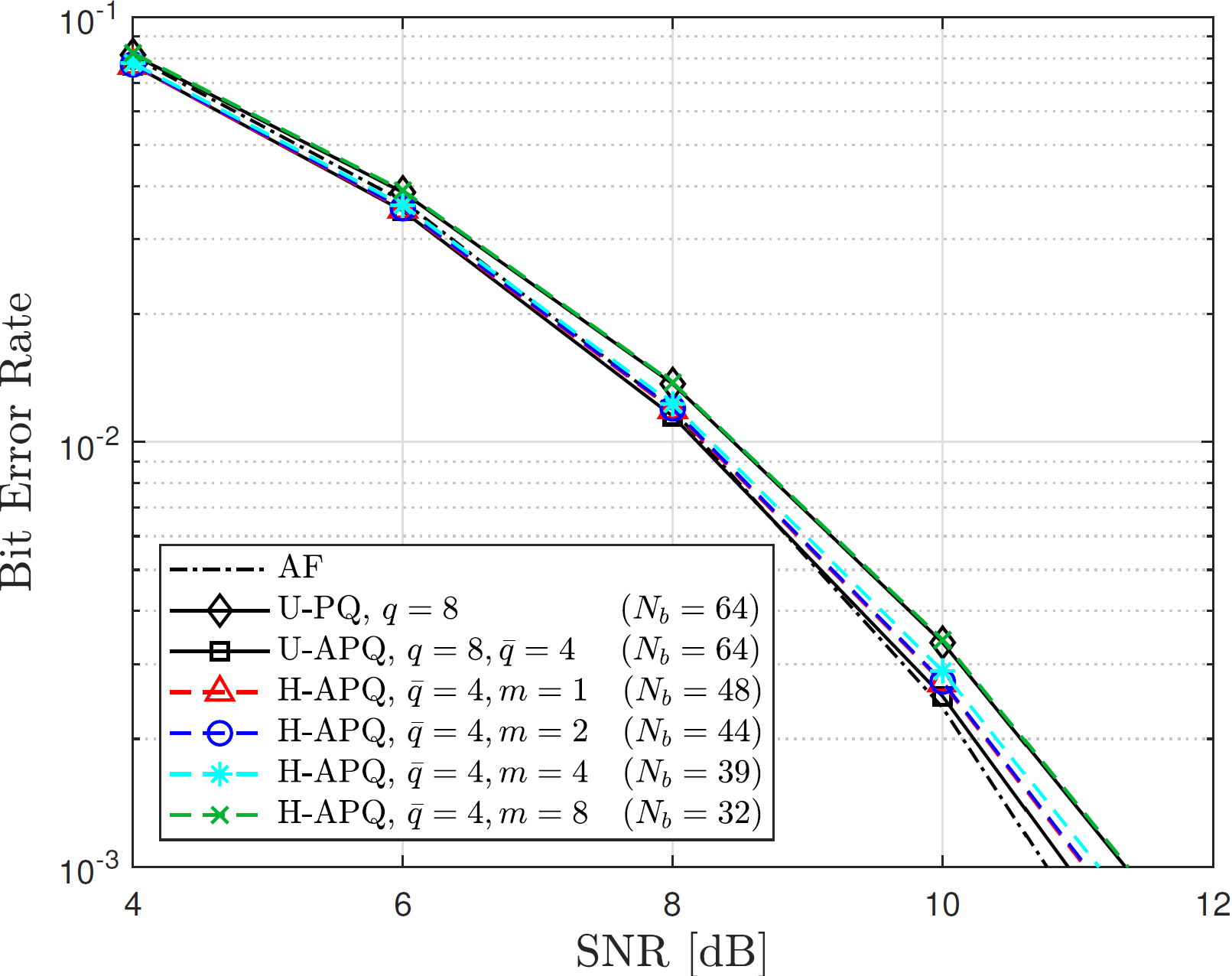}
        \caption{$N_{\rm S}=N_{\rm R}=N_{\rm D}=8$ }
        \label{fig:berNR8}
    \end{subfigure}\\[7pt]
    \begin{subfigure}{0.8\linewidth}
        \centering
        \includegraphics[width=\linewidth]{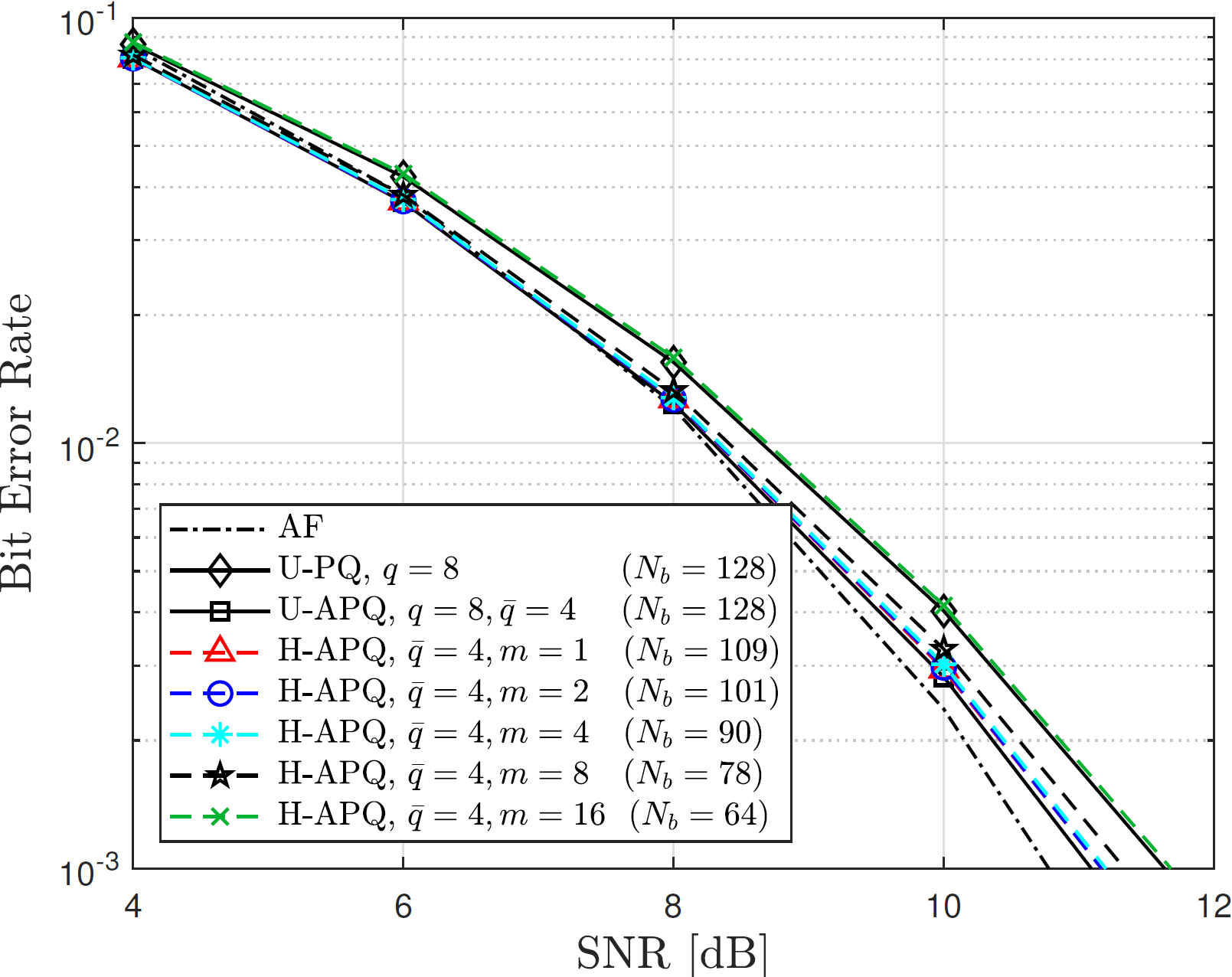}
        \caption{$N_{\rm S}=N_{\rm R}=N_{\rm D}=16$ }
        \label{fig:berNR16}
    \end{subfigure}\\[7pt]
    \caption{BER comparison for various quantization algorithms in the autoencoder-based MIMO QF relay systems with $M=4$, where   H-APQ applies $a_k= k\Delta$.}
    \label{fig:simulation_kdelta}
\end{figure}
\begin{figure}[t]
    \centering
    \begin{subfigure}{0.8\linewidth}
        \centering
        \includegraphics[width=\linewidth]{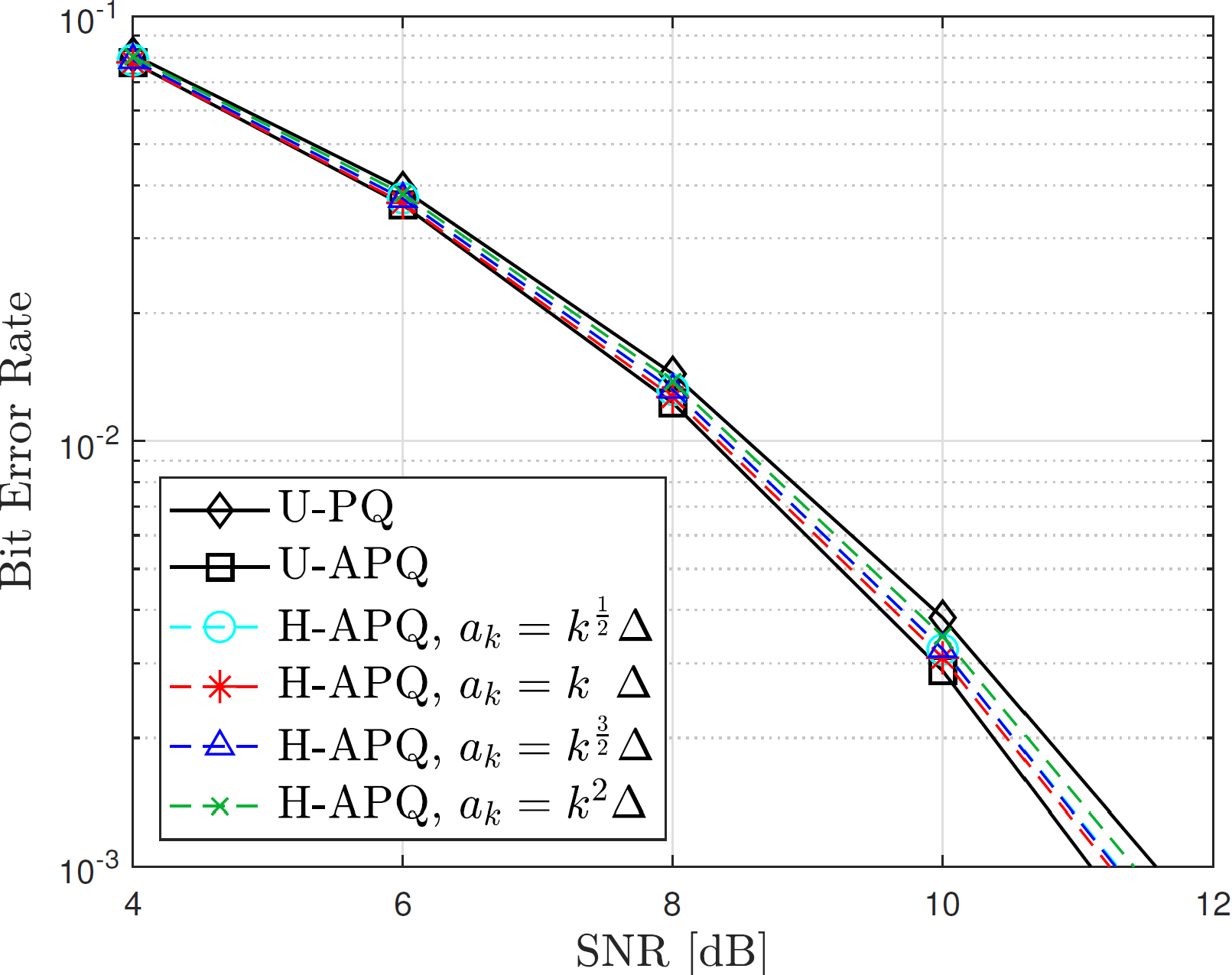}
        \caption{$m=1$ }
        \label{fig:berNR4m1}
    \end{subfigure} \\[7pt]
    \begin{subfigure}{0.8\linewidth}
        \centering
        \includegraphics[width=\linewidth]{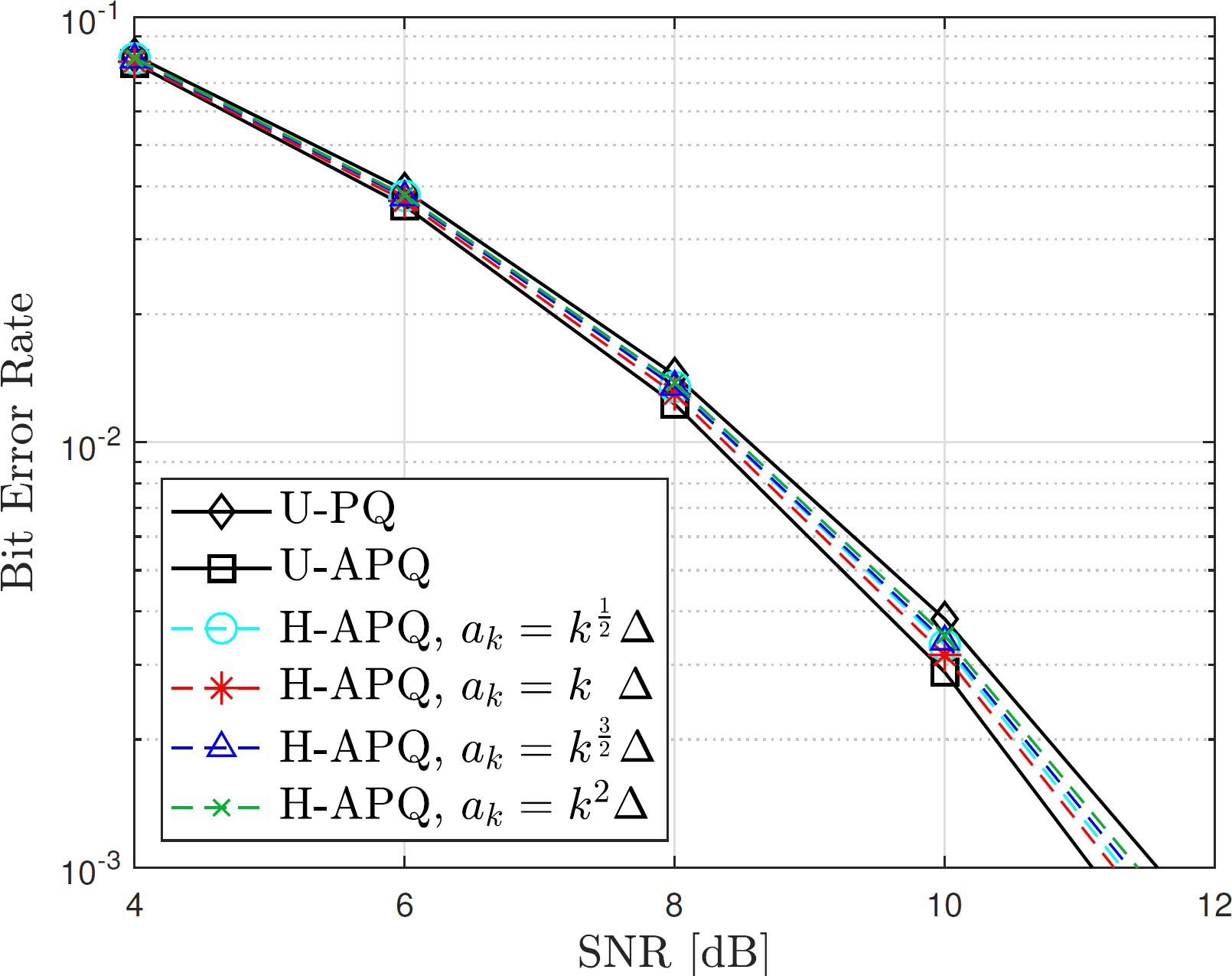}
        \caption{$m=2$}
        \label{fig:berNR4m2}
    \end{subfigure}\\[7pt]
    \caption{BER comparison   for  H-APQ   with various kinds of quantization levels, $a_k$,  when $N_{\rm S}\!=\!N_{\rm R}\!=\!N_{\rm D}\!=\!4$ and $M\!=\!4$.}
    \label{fig:hapq-ak}
\end{figure}
\section{Numerical Result}\label{sec:numerical_results}

In Section \ref{sec:sys_model}, we introduced the H-APQ algorithm as a memory-efficient quantization method for MIMO QF relay channels. 
This section compares its performance against existing quantization methods.

\subsection{Memory requirement at relay}

Since the comparison of required quantization bits in Table \ref{tab:bits} may not be sufficiently intuitive, we provide an illustration for better visualization.
Figure \ref{fig:memory_bit} illustrates that U-PQ with $q=8$ and U-APQ with $q=8, \bar{q}=4$ require the highest memory usage. 
In contrast, H-APQ significantly reduces memory consumption, particularly for larger values of $m$. 
Even when $m=1$ (denoted by  $\triangle$ markers),   H-APQ remains more memory-efficient than other methods for $N_{\rt}\le 20$.
This highlights how H-APQ effectively saves memory usage by adjusting the parameter $m$ unlike conventional quantization methods.

\subsection{System  performance}

The deep learning models are trained using the Adam optimizer \cite{Adam2014} within the TensorFlow framework \cite{tensorflow}. 
To evaluate the bit error rate (BER), we assume Rayleigh fading channels, i.e.,  ${\bf H}_{kl} \sim\CN(0,I_{N_l}), k \in\{ {\mathrm S}, {\mathrm R} \},l \in\{ {\mathrm R}, {\mathrm D} \}$, for both  training and test.
For $N_\st=4$, we use a training dataset of $N_{\rm train}=10^5$. 
As the model size increases, the dataset is expanded to $N_{\rm train}=2 \times 10^5$ for $N_\st=8$ and $N_{\rm train}=3 \times 10^5$  for $N_\st=16$ to accommodate the larger model.


%
\subsubsection{H-APQ with $a_k=k\Delta$}

Figure \ref{fig:simulation_kdelta} illustrates the BER performance of autoencoder-based MIMO QF relay systems with $M=4$ for different quantization methods where H-APQ employs $a_k= k\Delta$.
The  ``AF" curve represents the autoencoder-based MIMO AF relay system, serving as the performance benchmark.
The key observations are as follows:
\begin{itemize}
\item U-APQ with $q=8, \bar{q}=4$ achieves the best performance, closely approaching AF. 
However, this comes at the cost of high memory usage ($ N_b=32 $ for $N_{\rm R} = 4 $, $ N_b=64 $ for $ N_{\rm R} = 8 $, and $ N_b=128 $ for $N_{\rm R} = 16$).  
\item U-PQ with $ q=8 $, despite using the same number of quantization bits as U-APQ, performs the worst.  
\item H-APQ with $ m=N_{\rm R} $, equivalent to U-PQ with $ q=4 $, performs similarly to U-PQ with $ q=8 $, highlighting the performance saturation of U-PQ as the number of quantization bits increases.
\item H-APQ with $ m<N_{\rm R} $ achieves intermediate performance.
Notably, H-APQ with $m=2 $ offers near-optimal performance close to U-APQ while significantly reducing  memory usage, by up to 13 bits  for $ N_{\rm R}= 4$, 20 bits  for $ N_{\rm R}  = 8 $, and 27 bits  for $ N_{\rm R}  = 16$.
Moreover, as $ N_{\rm R}$   increases, the performance of H-APQ with $ m=2 $  progressively converges to that of U-APQ.

\end{itemize}
These findings demonstrate that H-APQ effectively balances performance and memory constraints.
Furthermore, by appropriately selecting $m$, the method can optimize performance while accommodating the relay's memory constraints.

\subsubsection{H-APQ with different  $a_k$ configurations}
As discussed in Section \ref{sec:H-APQ}, various quantization levels can be applied in the O-AQ method. 
Figure \ref{fig:hapq-ak} compares BER performance for $N_{\rm S}\!=\!N_{\rm R}\!=\!N_{\rm D}\!=\!4$ and $M\!=\!4$ when different quantization levels are used: $ a_k = k^{\frac{n}{2}} \Delta$  with $n = 1,2,3,4$.  
The ``U-PQ" and ``U-APQ" curves, corresponding to the cases  $q=8$ and $ q=8, \bar{q}=4 $, respectively,  serve as reference points for performance evaluation.
One can observe that:
\begin{itemize}
\item The BER curves for all cases fall between U-PQ ($ q=8$) and U-APQ ($ q=8, \bar{q}=4 $).  
\item $ a_k = k\Delta $ achieves the best performance, while $ a_k = k^2\Delta $ performs the worst.  
\item $ a_k = k^{\frac{1}{2}}\Delta $ and $ a_k = k^{\frac{3}{2}}\Delta $ exhibit similar intermediate performance.  
\end{itemize}
These results indicate that evenly spaced quantization levels, i.e., $ a_k = k\Delta $, yield excellent performance, although they may not always be optimal. 
Conversely, if spacing between quantization levels either decreases or increases with $ k $, distinguishing amplitude levels becomes more challenging, leading to degraded performance.  

Finally, modifying the quantization levels for O-AQ does not alter the number of required quantization bits. 
Thus, H-APQ with alternative quantization levels still offers significant memory savings over U-PQ and U-APQ, allowing for adaptive application based on system characteristics.

\section{Conclusion}\label{sec:conclusion}
This letter introduced the H-APQ algorithm, which integrates the O-AQ method for amplitude quantization, adaptively quantizing received relay signal amplitudes based on their relative magnitudes. 
While U-PQ experiences performance saturation even with increased quantization bits and U-APQ improves performance at the cost of high memory usage, H-APQ effectively balances these trade-offs. 
Additionally, we examined various quantization-level strategies for O-AQ and demonstrated that evenly spaced quantization levels can achieve strong performance. 
By significantly reducing memory requirements at the relay while maintaining competitive overall performance, H-APQ offers a practical and efficient solution for MIMO QF relay systems.
\ifCLASSOPTIONcaptionsoff
  \newpage
\fi


\begin{thebibliography}{1}
\bibitem{cooperative-diversity2004} J. N. Laneman, D. N. C. Tse and G. W. Wornell, “Cooperative diversity in wireless networks: Efficient protocols and outage behavior,'' {\em IEEE Trans. Inf. Theory}, vol. 50, no. 12, pp. 3062\,--\,3080, Dec. 2004.
\bibitem{relay-df2011} X. Jin, J.-S. No, and D.-J. Shin, “Relay selection for decode-and-forward cooperative network with multiple antennas,”
{\em IEEE Trans. Wireless Commun.}, vol. 10, no 12, pp. 4068\,--\,4079, Dec. 2011.
\bibitem{QF2008} M. Souryal and H. You, “Quantize-and-forward relaying with M-ary phase shift keying,” {\it 2008 IEEE Wireless Commun. and Netw. Conference, Las Vegas, NV.}, pp. 42\,--\,47, Apr. 2008.
\bibitem{JinQFtvt2022} X. Jin, ``Cooperative linear combining on quantize-forward relay channel with $M$-ary phase shift keying,'' {\it  IEEE Trans. Veh. Technol.}, vol. 71 no. 5, pp. 5645\,--\,5650, May 2022.
\bibitem{JinQFtwr2023} J. Park and X. Jin, ``Low-complexity linear signal detection on two-way quantize-forward relay channel," {\it IEEE Wireless Commun. Letters}, vol. 12, no. 12, pp. 2208\,--\,2212, Dec. 2023.
\bibitem{AQF2006}A. Steiner and S. S. Shamai, ``Single-user broadcasting protocols over a two-hop relay fading channel,'' {\it IEEE Trans.  Inf. Theory}, vol. 52, no. 11, pp. 4821\,--\,4838, Nov. 2006.

\bibitem{JinAQFaccess2022} X. Jin, ``Amplify-quantize-forward relay channel with quadrature amplitude modulation,'' {\it IEEE Access}, vol. 10, pp. 89445\,--\,89457, 2022.

\bibitem{JinMIMOAQF_TVT2024} X. Jin, ``MIMO detection on amplify-quantize-forward relay channel,'' {\it IEEE Trans. Veh. Technol.}, vol. 73, no. 6, pp. 8473\,--\,8486, Jun. 2024.
\bibitem{AE-MIMO-QF2024}J. Shin and X. Jin, ``Autoencoder–based MIMO cooperative communications with quantize--forward relaying," {\it IEEE Trans. Vehicular Technology.},  vol. 73, no. 12, pp. 19229\,--\,19239, Dec. 2024.
\bibitem{Adam2014} D. Kingma and J. Ba, ``Adam: A method for stochastic optimization,'' {\em arXiv preprint arXiv:1412.6980}, 2014.
\bibitem{tensorflow}M. Abadi, A. Agarwal, P. Barham, E. Brevdo, Z. Chen, C. Citro, G. S. Corrado, A. Davis, J. Dean, M. Devin, et al., 
``Tensorflow: Large-scale machine learning on heterogeneous distributed systems,'' {\em arXiv preprint},
2016. url  = {http://arxiv.org/abs/1603.04467}


\end{thebibliography}
\end{document}